\input harvmac
\noblackbox
\font\ticp=cmcsc10
 
\def\Title#1#2{\rightline{#1}\ifx\answ\bigans\nopagenumbers\pageno0\vskip1in
\else\pageno1\vskip.8in\fi \centerline{\titlefont #2}\vskip .5in}

\font\ticp=cmcsc10
\font\ttsmall=cmtt10 at 8pt

%

%
\def\({\left (}
\def\){\right )}
\def\[{\left [}
\def\]{\right ]}
\def\p{\partial}
\def\e{\epsilon}
\def\g{\gamma}

%
\lref\pol{J. Polchinski, ``Dirichlet-branes and Ramond-Ramond charges",
Phys. Rev. Lett. {\bf 75} (1995) 4724, hep-th/9510017.}
\lref\cvyo{M. Cvetic and D. Youm, ``General rotating five dimensional 
black holes
of toroidally compactified heterotic string", Nucl. Phys. 
{\bf B476} (1996) 118, hep-th/9603100.}
\lref\sen{A. Sen, ``Black hole solutions in heterotic string theory
on a torus", Nucl. Phys. {\bf B440} (1995) 421, hep-th/9411187.}
\lref\hopo{G.T. Horowitz and J. Polchinski, ``A correspondence principle
for black holes and strings", hep-th/9612146.}
\lref\mald{J.M. Maldacena, ``Statistical entropy of near extremal
five-branes", Nucl. Phys. {\bf B477} (1996) 168, hep-th/9605016.}
\lref\dkl{For a review, see M.J. Duff, R.R. Khuri, and J.X. Lu, ``String
Solitons", Phys. Rep. {\bf 259} (1995) 213, hep-th/9412184.}
\lref\pen{R. Penrose, ``Gravitational collapse: The role of general
relativity", Rev. del Nuovo Cimento, {\bf 1} (1969) 252.}
\lref\ascv{A. Strominger and C. Vafa, ``Microscopic origin of the
Bekenstein-Hawking entropy'', Phys. Lett. {\bf B379} (1996) 99,
hep-th/9601029.}
\lref\ghrev{G.T. Horowitz, ``The origin of black holes in string
theory'', gr-qc/9604051. }
\lref\jmthes{J.M. Maldacena, ``Black holes in string theory'',
Ph.D. thesis, hep-th/9607235.}
\lref\hms{ G.T. Horowitz, J.M. Maldacena and A. Strominger, ``Nonextremal
black hole microstates and U-duality'', Phys. Lett. {\bf B383} (1996)
151, hep-th/9603109.}
\lref\dm{S.R. Das and S.D. Mathur, ``Excitations of D-strings, entropy
and duality'', Phys. Lett. {\bf 375B} (1996) 103, hep-th/9601152.}
\lref\dmw{A. Dhar, G. Mandal and S.R. Wadia, ``Absorption vs. decay of
black holes in string theory and T-symmetry'', Phys. Lett. {\bf 388B}
(1996) 51, hep-th/9605234.} 
\lref\dam{S.R. Das and S.D. Mathur. ``Comparing
decay rates for black holes and D-branes'', Nucl. Phys. {\bf B478}
(1996) 561, hep-th/9606185.}
\lref\ms{J.M. Maldacena and A. Strominger, ``Black hole greybody factors
and D-brane spectroscopy'', Phys. Rev. D {\bf 55} (1997) 861,
hep-th/9609026.} 
\lref\km{I.R. Klebanov and S.D. Mathur, ``Black hole greybody factors
and absorption of scalars by effective strings'', hep-th/9701187.}
\lref\dol{S.W. Hawking, ``The unpredictability of quantum gravity'',
Commun. Math. Phys. {\bf 87} (1982) 395.}
\lref\hbh{S.W. Hawking, ``Particle creation by black holes'',
Commun. Math. Phys. {\bf 43} (1975) 199.}  
\lref\sussug{L. Susskind and J. Uglum, ``String physics and black holes'',
Nucl. Phys. Proc. Suppl. {\bf 45BC} (1996) 115,  hep-th/9511227, and
references therein.} 
\lref\dbh{D. Garfinkle, G.T. Horowitz and A. Strominger, ``Charged
black holes in string theory'', Phys. Rev. D {\bf 43} (1991) 3140,
Erratum {\it ibid.} {\bf 45} (1992) 3888.} 
\lref\gm{G.W. Gibbons, ``Antigravitating black hole solitons with
scalar hair in $N=4$ supergravity'', Nucl. Phys. {\bf B207} (1982)
337; G.W. Gibbons and K. Maeda, ``Black holes and membranes in higher
dimensional theories with dilaton fields'',{\it ibid.} {\bf B298}
(1988) 741.}
\lref\klopp{R. Kallosh, A. Linde, T. Ort\'{\i}n, A.
Peet, and A.  Van Proeyen, ``Supersymmetry as a cosmic censor'',
Phys. Rev. D {\bf 46} (1992) 5278.} 
\lref\pbr{G.T. Horowitz and A. Strominger, ``Black strings and
$p$-branes'', Nucl. Phys. {\bf B360} (1991) 197.}
\lref\hs{G.T. Horowitz and A.R. Steif, ``Strings in strong
gravitational fields'', Phys. Rev. D {\bf 42} (1990) 1950.}
\lref\ght{G.W. Gibbons, G.T. Horowitz and P.K. Townsend, ``Higher
dimensional resolution of dilatonic black hole singularities'',
Class. Quant. Grav. {\bf 12} (1995) 297, hep-th/9410073.}
\lref\holwil{C.F.E. Holzhey and F. Wilczek, ``Black holes as
elementary particles'', Nucl. Phys. {\bf B380} (1992) 447,
hep-th/9202014.}

%
\baselineskip 12pt
\Title{\vbox{\baselineskip12pt
\line{\hfil  UCSBTH-97-05}
\line{\hfil \tt hep-th/9704058} }}
{\vbox{
{\centerline{Naked Black Holes }}
}}
\centerline{\ticp Gary T. Horowitz\footnote{}{\ttsmall
gary@cosmic.physics.ucsb.edu, sross@cosmic.physics.ucsb.edu}
and Simon F. Ross }
\bigskip
\vskip.1in
\centerline{\it Department of Physics, University of California,
Santa Barbara, CA 93106, USA}
\bigskip
\centerline{\bf Abstract}
It is shown that there are large static black holes for which all
curvature invariants are small near the event horizon, yet any object
which falls in experiences enormous tidal forces {\it outside} the
horizon. These black holes are charged and near extremality, and exist
in a wide class of theories including string theory.  The implications
for cosmic censorship and the black hole information puzzle are
discussed.

\Date{April, 1997}

\baselineskip=16pt
\newsec{Introduction}

It is commonly believed that the spacetime curvature is small near the
horizon of a large static black hole, and objects can fall in without
being disrupted. We will show that this is not always the case. There
are black holes in which the area of the event horizon is large and
all curvature invariants are small near the horizon. Nevertheless, any
object which falls in experiences enormous tidal forces outside the
horizon.  This is a result of the fact that the curvature is actually
very large and almost null near the horizon. When measured in a static
frame (which is also becoming null), the components of the curvature
remain small. This implies that all curvature invariants are small,
and perturbative $\alpha'$ corrections in string theory are
negligible.  However, in a freely falling frame, the curvature
components are very large. Since the region of large tidal forces is
visible to distant observers, we will call such objects ``naked black
holes".

These black holes exist in a wide class of theories including the
supergravity theories that arise in the low energy limit of string
theory. Our examples are all charged black holes which are either at
or near extremality. In fact, many of the solutions to string theory
which have been found in recent years \dkl\ (including the higher
dimensional black $p$-branes) have limits in which they become naked
in the above sense. Some of these solutions are known to have the
property that the horizon shrinks down to zero size and becomes
singular as one approaches the extremal limit (for fixed mass). We
will show that the curvature felt by an infalling observer can become
large even when the area of the horizon remains large.

In retrospect, it is not surprising that there can be a significant
difference between the size of the curvature seen by static and freely
falling observers. After all, static observers measure the curvature
in the rest frame of the black hole. Near the horizon, freely falling
observers are highly boosted with respect to the static ones, and one
would expect their curvature components to be much larger. From this
viewpoint, it is surprising that the familiar Schwarzschild and
Reissner-Nordstr\"om black holes have approximately the same size
curvature in the static and freely falling frames. This is possible
only because of a special cancellation between different components of
the curvature in these metrics: Their curvature is actually invariant
under radial boosts. Based only on these two examples, one might have
concluded that boost invariance of the curvature was somehow implied
by the structure of the event horizon. We will see that this is not
the case.

The existence of naked black holes has implications for cosmic
censorship \pen. Although they do not affect a strict interpretation
of this conjecture in terms of singularities, they weaken the spirit
of cosmic censorship. One of the motivations for this conjecture was
to show that general relativity would break down only in regions of
spacetime shielded by event horizons. While naked black holes have
nonsingular event horizons, the curvature outside can be larger than
the Planck scale. Thus, effects of quantum gravity could be visible
outside macroscopic black holes. Although we will consider the eternal
static black holes, it seems likely that one can form the near
extremal black holes from regular initial data.

Naked black holes may also play a role in resolving the black hole
information puzzle. Recently, the entropy of certain extremal and near
extremal black holes has been reproduced in string theory by counting
states in the limit of weak coupling
\refs{\ascv,\ghrev,\jmthes}. Furthermore, the radiation 
produced at weak coupling has the same spectrum as the Hawking
radiation emitted by the black hole \refs{\dam,\ms}.  However, the
weak coupling description is manifestly unitary, while Hawking has
given arguments that black hole radiation is not unitary
\refs{\dol,\hbh}. Of course a basic assumption in his arguments is
that matter can fall into a large black hole undisturbed, which makes
it difficult to see how the information about the state of the matter
gets out. We will see that some of the near extremal black holes whose
entropy has been understood by counting string states are, in fact,
naked. Matter falling into such black holes will be disrupted by the
large tidal forces, which may invalidate the simple argument for
information loss in these cases.

In the next section, we discuss the difference between the curvature
in a static frame and in an infalling frame for a general class of
metrics.  Section 3 contains some examples of naked black holes in
general relativity and section 4 contains examples from string
theory. Some implications of the existence of these black holes are
discussed in section 5.

\newsec{Curvature and Tidal Forces}

We begin by considering  the following class of metrics
in $d$ spacetime dimensions,  
\eqn\gmet{ds^2 = -{F(r) \over G(r)} dt^2  + {dr^2 \over
F(r)} + R^2(r) d\Omega_{n+1}+ H^2(r) dy^i dy_i,}
where $i=n+3,\ldots,d-1$. This class includes most of the recently
discussed black hole and black $p$-brane ($p=d-n-3$) solutions; the
metric will have a horizon at $r=r_0$ if $F(r_0)=0$.  The curvature
is, of course, completely characterized by the components of the
Riemann tensor in an orthonormal frame. Let us consider first the
static frame
\eqn\sframe{\eqalign{(e_0)_\mu = &-F^{1/2}(r)G^{-1/2}(r)\ \p_\mu t,
\qquad (e_1)_\mu = 
F^{-1/2}(r)\ \p_\mu r, \cr (e_a)_\mu =& R(r) \sin\theta_1 \ldots
\sin\theta_{a-2}\ \p_\mu\theta_{a-1}, \cr (e_{n+2})_\mu =& R(r)
\sin\theta_1\ldots \sin\theta_n\ \p_\mu \phi, \cr (e_i)_\mu =&
H(r)\ \p_\mu y_i,}} 
where $\theta_{a-1}$, $a=2, \ldots, n+1$ and $\phi$ are coordinates on
$S^{n+1}$. The only non-vanishing components of the curvature in this
orthonormal frame are $R_{0101}$, $R_{0k0k}$ (no sum on $k$),
$R_{1k1k}$, and $R_{klkl}$, where $k, l = 2, \ldots, d-1$ (and
components related to these by symmetry). As we will see below,
radially infalling observers will measure the curvature not in this
static frame, but in terms of another orthonormal frame related to
\sframe\ by a local radial boost. That is, a frame where
\eqn\boostframe{(e_{0'})_\mu = \cosh \alpha (e_0)_\mu + \sinh \alpha
(e_1)_\mu, \qquad (e_{1'})_\mu = \sinh \alpha (e_0)_\mu + \cosh \alpha
(e_1)_\mu, } 
the other $(e_k)_\mu$ are as before, and $\alpha = \alpha(x^\mu)$ is
some function of the coordinates. 

The components of the curvature in any boosted frame cannot be smaller
than the components in the static frame. This can be seen as follows.
The non-vanishing components of the curvature in the boosted frame
will be
\eqn\bcurv{\eqalign{R_{0'1'0'1'} &= R_{0101},\qquad  R_{0'k1'k} = \cosh
\alpha \sinh \alpha (R_{0k0k} +R_{1k1k}),
\cr R_{0'k0'k} &= 
 R_{0k0k} + \sinh^2 \alpha (R_{0k0k} + R_{1k1k}),
\cr R_{1'k1'k} &=  
 R_{1k1k} + \sinh^2 \alpha (R_{0k0k} + R_{1k1k}),}}
and $R_{klkl}$. For each $k$, consider the larger of the two 
components $R_{0k0k}, \
R_{1k1k}$.  Since $R_{0k0k} + R_{1k1k}$ has the same sign as the
larger component, it is clear that its magnitude cannot decrease under
a boost. The curvature can remain unchanged if $R_{0k0k} = -
R_{1k1k}$; this occurs, for example, in the Schwarzschild solution. In
general, the boosted components are larger and the curvature is
minimized in the static frame.  (The static frame is preferred because
it is the only one for which the $R_{0k1k}$ components vanish.)  The
static frame is thus the most convenient one for calculating curvature
invariants and contractions of the Riemann tensor. (In the examples we
consider, the size of $R_{0k0k}$, $R_{1k1k}$ in the static frame is
comparable to that of $R_{0101}$, $R_{klkl}$.)

To determine the physical effect of the curvature on geodesic
observers, we need to use an orthonormal frame which is parallelly
propagated along the geodesics. If we consider timelike geodesics in
the metric \gmet, with proper time $\tau$ and tangent vector $u^\mu =
dx^\mu/d\tau$, we can choose coordinates on $S^{n+1}$ so that
the geodesic lies in the equatorial plane. There are a number of
constants of motion:
\eqn\consts{E = {F(r) \over G(r)} \dot{t}, \qquad p_i = H^2(r)
\dot{y}_i, \qquad p_\phi = R^2(r) \dot{\phi},}
where a dot denotes $d/d\tau$. For the sake of simplicity, we will
consider radial geodesics, $p_i = p_\phi =0$. From the normalization
condition $u^\mu u_\mu= -1$, we can see that
\eqn\rdot{\dot{r}^2 =  E^2 G(r) - F(r).} 
The parallelly propagated orthonormal frame, in which $(e_{0'})_\mu =
u_\mu$, is then related to the static frame by a radial boost,
\eqn\inframea{\eqalign{(e_{0'})_\mu &= u_\mu = -E \p_\mu t + {\dot{r}
\over F(r)} \p_\mu r \cr  &= \cosh \alpha
(e_0)_\mu + \sinh \alpha (e_1)_\mu,}}
and 
\eqn\inframeb{(e_{1'})_\mu = \sinh \alpha (e_0)_\mu + \cosh \alpha
(e_1)_\mu, } 
where $\cosh \alpha = E [G(r)/F(r)]^{1/2}$. Note that since the
horizon lies at $F(r) = 0$, the boost parameter $\alpha$ diverges as
we approach the horizon.

We can compute the components of the
curvature in this frame by first computing the components in the
static frame, and then applying the transformations \bcurv. However,
there is another, simpler route to computing the boosted components,
which offers a more direct physical understanding. We can see from
\bcurv\ that the difference between the static frame and the boosted
frame is essentially the same for all components, so it suffices to
calculate $R_{0'k0'k}$. These components correspond to tidal forces in
the transverse directions. In other words, they measure the relative
acceleration of nearby geodesics. In fact, they are simply given by
\eqn\result
{R_{0'a0'a} = -{\ddot R\over R}, \qquad \qquad R_{0'i0'i} = -{\ddot
H\over H},}
as we now show. For a family of radial infalling geodesics with
tangent vector $u^\mu$, and the set of deviation vectors $\eta =
\p/\p\theta_{a-1}$ for $a=2,\ldots, n+1$, and $\eta = \p/\p \phi$ for
$a=n+2$, we have
\eqn\etaa{u^\nu \nabla_\nu \eta^\sigma = u^\nu \Gamma_{\nu\rho}^\sigma 
\eta^\rho = u^r {R' \over R} \eta^\sigma = {\dot R \over R}
\eta^\sigma.} 
The geodesic deviation equation then implies
\eqn\gdeva{R_{\mu\nu\rho}\,^\sigma
u^\mu  \eta^\nu u^\rho  = - u^\mu \nabla_\mu (u^\nu \nabla_\nu
\eta^\sigma) = -{\ddot R \over R}\eta^\sigma.}
Thus
\eqn\gdevb{\eqalign{R_{0'a0'a} &= R_{\mu\nu\rho}\,^\sigma u^\mu
(e_a)^\nu u^\rho (e_a)_\sigma = -{\ddot{R}\over R} = -{1 \over R}(R''
\dot{r}^2 + R' 
\ddot{r})\cr & =  -{1 \over R}\left[ R'' \left( E^2 G - F
\right) + {R' \over 2} \left( E^2 G'  - F' \right) \right],}} 
where we have used \rdot. Similarly, $R_{0'i0'i} = -\ddot{H}/H$ for $i
= n+3,\ldots,d-1$. The terms proportional to $E^2$ in this expression
correspond to the enhancement of the curvature in the geodesic frame
over the static frame.  Note that although the boost parameter
diverges at the horizon, these terms will generally be finite. This is
due to a cancellation between the leading order contributions to
$R_{0k0k}$ and $R_{1k1k}$ near the horizon. For Schwarzschild, there
is complete cancellation; the terms proportional to $E^2$ vanish
because $R'' = G' = 0$. It is clear that whenever these terms do not
vanish, the tidal force can be made arbitrarily large, simply by
taking the conserved energy per unit mass along the geodesic $E$ to be
large. But this is also true for nonradial geodesics in
Schwarzschild. Conversely, no matter how large the tidal force is, we
can find a family of geodesics for which its effect is small simply by
decreasing $E$. To avoid this ambiguity, we will always assume the
energy per unit mass $E$ is of order one, i.e. we will consider
geodesics that start at infinity with small velocity.  When we
calculate $R_{0'k0'k}$ in the examples, we will only keep the part
proportional to $E^2$, which represents the difference between the
static frame and the boosted frame.

\newsec{Examples from General Relativity}

In this section we discuss two examples of black holes with large
horizon area and small curvature components in the static frame, but
large curvature in a freely falling frame. Both examples are four
dimensional black holes in general relativity coupled to gauge fields
and scalar fields.  The solutions will take the familiar form
\eqn\gensol{ds^2 = -F(r) dt^2 + {dr^2 \over F(r)} + R^2(r)
d\Omega_2.}
For metrics of this type, the difference between the curvature in the
static and infalling frames can be easily seen as follows.  The event
horizon is at $r=r_+$, the largest value of $r$ for which $F$
vanishes, and it has area $4\pi R(r_+)^2$. The static curvature
near the horizon contains the components
\eqn\stcur{R_{2323} = {1 \over R^2}, \qquad
R_{0202}={F'R'\over 2R}.} 
However, the requirement of positive energy density $G_{00} \ge 0$,
implies that, near the horizon, $1/R^2 \ge F'R'/R$. So the static
curvature will be small (in Planck units) if $R(r_+) \gg1$.  The
curvature in a freely falling frame is given by \gdevb\ with $G(r) =1$,
\eqn\bhtide{R_{0'20'2} = -{1 \over R}\left[ R''(E^2 - F) - {F'
R' \over 2}\right]. }
Near the horizon, $F(r)$ is small, so this will be larger than the
Planck scale if $|R''/R| >1$. Thus to satisfy our conditions, we
simply require $R\gg1$ and $|R''/R|>1$ at $r=r_+$. In the examples
below, the tidal forces are of the same magnitude over a radial proper
distance of   order $R(r_+)$ outside the horizon. Thus, $R(r_+)$
is a measure of the size of the region of large tidal forces.

\subsec{Dilaton Black Holes}

Our first example is the dilaton black hole metrics
\refs{\gm,\dbh}. These are solutions of a theory with a single Maxwell
field $F_{\mu\nu}$ and scalar field $\phi$ with the coupling between
the Maxwell field and the scalar field governed by an arbitrary
constant $a$. The action is
\eqn\dbhac{S = \int d^4 x \sqrt{-g} \[R - 2 (\nabla
\phi)^2 - e^{-2a\phi} F_{\mu\nu}F^{\mu\nu}\],}
and the metric for a dilaton black hole  is given by
\gensol\ with
\eqn\uuf{F(r) = {(r-r_+)(r-r_-) \over R^2}}
and
\eqn\R{R(r) = r \left( 1- {r_- \over r} \right)^{a^2/(1+a^2)}.}
There is a horizon at $r=r_+$ and a singularity at $r=r_-$ for $a \neq
0$. For $a=0$, this metric reduces to the Reissner-Nordstr\"om metric;
$r=r_-$ is an inner horizon, and there is a singularity at $r=0$.  The
extremal limit in both cases is $r_+ = r_-$.  The ADM mass and charge
are
\eqn\dbhmass{\eqalign{
M&={r_+\over2} + \({1-a^2\over 1+a^2}\) {r_-\over 2},\cr
Q&=\({r_+ r_-\over 1+a^2}\)^{1/2}. }}

The horizon area will be large and the static curvature will be small
(in Planck units) if
\eqn\dbhr{R(r_+) = r_+\epsilon^{a^2/(1+a^2)} \gg 1,}
where $\epsilon\equiv (1-r_-/r_+)$.  Note that the exponent of
$\epsilon$ is always less than one.  As discussed above, the tidal
forces in the geodesic frame will be larger than the Planck scale if
\eqn\dbhtide{ \left|{R'' \over R} \right|
= {a^2 \over (1+a^2)^2} {(1-\epsilon)^2 \over r_+^2
\epsilon^2}>1. } 
This will be satisfied, for $a \neq 0$, if $r_+ \epsilon \ll 1$. Thus
we see that there is a range of parameters for which the curvature in
the static frame is small, but infalling observers experience large
tidal forces near the horizon, namely $\epsilon \ll 1$ and
$\epsilon^{-a^2/(1+a^2)} \ll r_+ \ll \epsilon^{-1}$. Physically, the
reason for the difference in the size of the curvature components is
that the curvature is becoming large and nearly null near the
horizon. Since the static frame is becoming null, it does not see this
effect. Since $\epsilon$ is small, these black holes are all close to
extremality, and since $r_+$ is large, they have a large mass. For
fixed mass, the area of the event horizon goes to zero in the extremal
limit. The spacetime develops a null singularity if $0<a \le 1$ and a
timelike singularity if $a>1$.  We are considering a different limit
in which the mass is increased as one approaches extremality, so the
horizon area remains large.

One might wonder whether infalling observers will experience large
tidal forces only in cases where the extremal limit is singular.  This
seems to be suggested by the above example, since the one case where
the extremal limit is nonsingular, the Reissner-Nordstr\"om metric
($a=0$), is also the one case where the tidal forces remain small for
infalling observers. However, this is not the case, as the next
example shows.

\subsec{$U(1)^2$ black holes}

Our second example is a class of black hole solutions to a theory with
two Maxwell fields and a scalar field \refs{\gm,\klopp}. The action is
\eqn\uact{S = \int d^4 x \sqrt{-g} \[R - 2 (\nabla
\phi)^2 - e^{-2\phi} (F_{\mu\nu} F^{\mu\nu}+G_{\mu\nu} G^{\mu\nu})\],}
and the metric for the black hole solutions is again given by
\gensol\ with 
\eqn\uf{F(r) = {(r-r_+)(r-r_-) \over R^2},}
but now
\eqn\ur{R^2(r) = r^2 - \Sigma^2.}
This metric has an event horizon at $r=r_+$, an inner horizon at
$r= r_-$, and a singularity at $r=|\Sigma|$.  One Maxwell field has an
electric charge $Q$ and the other has a magnetic charge $P$. In terms
of the mass and charges,
\eqn\mcff{\eqalign{
\Sigma &= {P^2 - Q^2 \over 2M}, \cr
r_\pm &= M \pm (M^2 +\Sigma^2 - Q^2 -P^2)^{1/2}.}}
If $Q=P$, then $\Sigma = 0$ and this metric is just the
Reissner-Nordstr\"om metric. If one of the charges vanishes, then
$|\Sigma| = r_-$ and it reduces to the dilaton black hole metric
discussed in the previous section with $a=1$. With both charges
nonzero, $|\Sigma| < r_-$, and there is a smooth horizon in the
extremal limit $r_+ = r_-$.

As before, the horizon area will be large and the static curvature
small if
\eqn\scurs{R(r_+) = r_+ \delta^{1/2} \gg1,}
where $\delta \equiv (1-\Sigma^2/r_+^2)$.  The curvature in the
geodesic frame will be larger than the Planck scale if
\eqn\utide{\left|{R'' \over R} \right|=
{ (1-\delta) \over r_+^2 \delta^2} >1, } 
at $r=r_+$. There is again a range of parameters for which both these
conditions are satisfied, namely $\delta \ll r_+^2 \delta^2 \ll
1$. Thus, even though the area of the event horizon is large even at
extremality, observers experience large tidal forces outside the black
hole.

Since $|\Sigma| < r_-$, $(1-r_-/r_+) < \delta$, and thus the above
condition implies that the black hole must be near-extreme for this
effect to appear. Further, since both $\Sigma$ and $r_-$ are close to
$r_+$, they must be close to each other.  This implies that one charge
is much greater than the other, so away from the horizon, this
solution resembles the $a=1$ dilaton black hole discussed above.

\newsec{Examples from String Theory}

In this section, we will consider black hole and black $p$-brane
solutions which arise in string theory.  We will consider both the
Einstein and the string metrics and show that in both cases, the tidal
forces experienced by infalling objects can be large when the horizon
area is large.  The metrics we discuss are solutions of the low energy
equations of motion, to leading order in $\alpha'$.  Since we are
interested in situations where the components of the curvature in a
freely falling frame become large, one might worry that $\alpha'$
corrections will become important. This is not the case.  The
equations of motion take the general form
\eqn\alphap{R_{\mu\nu} +{ \rm matter\ contributions} +
\alpha' R_{\mu}\,^{\rho\sigma\lambda}
R_{\nu\rho\sigma\lambda} + \ldots = 0.}
We can compare the size of the $\alpha'$ correction to the leading
order term in any frame we choose, since under a boost, both
quantities will be equally boosted. In particular, if the curvature in
the static frame is small, so that the curvature squared term is small
compared to the first term, then in the geodesic frame, there will be
cancellations in the calculation of the $\alpha'$ term which make it
the same relative size as in the static frame. We will see that one
can also choose the parameters so that $g e^{\phi}$ is small in the
region where the tidal forces become large, so perturbative quantum
corrections should be small as well.

In the remainder of this section, we will set $\alpha'=1$, so we are
working in string units.

\subsec{Neveu-Schwarz charged black holes}

We first consider the black hole solution with electric
Neveu-Schwarz charges associated with internal momentum and string
winding number. The string metric is \sen
\eqn\nsmet{ds^2 = - \Delta^{-1} \(1-{r_0\over r}\) dt^2 + 
\(1-{r_0\over r}\)^{-1}
dr^2 + r^2 d\Omega_2,} 
where 
\eqn\dd{\Delta = \left( 1+ {r_0 \sinh^2 \gamma_1 \over r} \right) \left(
1 + {r_0 \sinh^2 \gamma_p \over r} \right)}
and the dilaton is given by
\eqn\nsdil{e^{2 \phi} = \Delta^{-1/2}.}
The ADM mass of these black holes is
\eqn\nsmass{M = {r_0 RV \over g^2}(2+\cosh 2\gamma_1 + \cosh
2\gamma_p),}
and the integer normalized charges are 
\eqn\nscharge{n = {R^2 V \over g^2} r_0 \sinh 2 \gamma_p, \qquad m =
{V \over g^2} r_0 \sinh 2 \gamma_1,}
where $R$ is the radius of a compact internal direction, and 
$(2\pi)^5 V$ is the volume of an internal five-torus. (We are using
the same conventions  as \jmthes.)

The curvature in the static frame is of order $1/r_0^2$ at the horizon
$r=r_0$, so we must take $r_0 \gg 1$ to keep it small. The
curvature in the infalling frame can be found from \gdevb\ with
$F(r) = (1-r_0/r)$, $G(r) = \Delta$, and $R(r) =r$. At $r=r_0$, it is
\eqn\nsgeod{R_{0'20'2} = -{R' \over 2 R} \left( G' E^2 - F'
\right) = -{E^2 \Delta' \over 2 r_0} + {1 \over 2 r_0^2}.}
Since 
\eqn\dd{\Delta'(r_0) = -{1 \over r_0}(\sinh^2 \g_1 \cosh^2 \g_p +
 \cosh^2 \g_1 \sinh^2 \g_p ),} 
we can make the tidal forces arbitrarily large by increasing $\g_1$
or $\g_p$.  Physically, this just corresponds to increasing the mass
and charges.

In the regime where the tidal forces are large, $e^\phi$ is very
small, and thus perturbative
quantum corrections are negligible. String $\alpha'$
corrections associated with powers of the curvature are also
negligible since the static curvature is small.  Since the dilaton
$\phi$ is large and negative, one might worry about $\alpha'$
corrections involving derivatives of the dilaton.  However, near
$r=r_0$,
\eqn\nphi{\partial_r \phi = -{1 \over 4} {\Delta' \over \Delta} \sim
{1 \over r_0} }
when $\g_1$ or $\g_p$ is large. Thus, $\alpha'$
corrections involving the dilaton are also unimportant. 

The extremal limit for this class of black holes is $r_0 \to 0$,
$\gamma_1, \gamma_p \to\infty$ with $n, m$ fixed. It may appear that
the large tidal forces are present far from the extremal limit, since
we have taken $r_0 \gg 1$. However, for fixed charges, the mass above
extremality is 
\eqn\dele{\Delta M = M - M_{ext} \approx {2 r_0 RV \over g^2}.}  
When $r_0 \gg 1$, $\Delta M$ is large, but $\Delta M/M$ is still small
since $\g_1$ or $\g_p$ is large.

The Einstein metric is obtained by multiplying \nsmet\ by
$e^{-2\phi}$.  The area of the horizon is thus increased by the factor
$\cosh\g_1 \cosh\g_p$ and the static frame curvature near the horizon
is decreased by the same factor. The Einstein metric takes the form
\gensol\ and one can verify that the tidal forces in this metric are
proportional to $1/r_0$. Thus, for the range of parameters we have
been considering, the tidal forces would not be large for infalling
observers. However, if one takes $r_0$ small and $\g_1,\g_p$
sufficiently large, then both the size of the black hole and the tidal
forces in the Einstein metric will be large. This example is closely
related to the examples of the previous section. If $\g_1 = \g_p$, the
Einstein metric is the same as the black hole discussed in section
{\it 3.1} with $a=1$ (or the one discussed in {\it 3.2} with
$P=0$). If $\g_1=0$ or $\g_p=0$, the metric is the same as the one in
{\it 3.1} with $a=\sqrt 3$.

\subsec{$p$-branes with a Ramond-Ramond charge}

The effect we have been discussing is also present for extended
objects in string theory. We now consider black $p$-branes with a
single Ramond-Ramond charge in $d=10$ \pbr. The string metric is
\eqn\pbmet{ds^2 = f^{-1/2} \left[ - \left( 1- {r_0^n \over r^n}
\right)dt^2 + dy^i dy_i \right] + f^{1/2} \left[ \left( 1- {r_0^n
\over r^n} \right)^{-1} dr^2 + r^2 d\Omega_{n+1} \right],}
where there are $p=7-n$ coordinates $y^i$, and
\eqn\pbf{f = 1 + {r_0^n \sinh^2 \alpha \over r^n}.}
The dilaton is given by $e^{2\phi} = f^{(n-4)/2}$. The charge is
\eqn\pbq{Q \sim {r_0^n \over g} \sinh 2\alpha.}
If we assume that the longitudinal coordinates $y_i$ are periodically
identified, so the $p$-brane is wrapped around a torus with volume
$V$, the mass is
\eqn\pbe{ M \sim {r_0^n V \over g^2} \left( {n+2 \over n} +
\cosh 2 \alpha \right),} 

The curvature in the static frame is of order $ 1/(r_0^2 \cosh
\alpha)$ at the horizon $r = r_0$. The curvature in the geodesic frame
can again be found from \gdevb\ with $F(r)=f^{-1/2}(1-r_0^n/r^n), G(r)
= 1$, $R(r) = f^{1/4} r$ and $H(r) = f^{-1/4}$
\eqn\pbgeoda{\eqalign{R_{0'a0'a} &=  -{1 \over R}\left[ R'' \left( E^2
- F \right) - {R' F'\over 2} \right] = {E^2 \over 4f^2} \({3 \over 4} f'^2
- f f'' - {2 \over r} ff'\) - \ldots \cr &= {E^2 \over 16 \tilde{f}^2
r^2}(4n - n^2) + \ldots}} 
for $a=2,\ldots,n+2$, and 
\eqn\pbgeodb{R_{0'i0'i} = -{1 \over H}\left[ H'' \left( E^2  - F
\right) - {H' F'\over 2} \right] = {E^2 \over 4f^2} \(-{5 \over 4} f'^2
+ f f''\) + \ldots = {E^2 \over 16\tilde{f}^2 r^2}(4n
- n^2) + \ldots} 
for $i = n+3, \ldots, 9$, where 
\eqn\pbtf{\tilde{f} = 1 + {r^n \over r_0^n \sinh^2 \alpha}.}
and we have kept only the terms proportional to $E^2$ on the right hand
side.
Notice that the leading order behaviors of $R_{0'a0'a}$ and $R_{0'i0'i}$
are exactly the same.

Let us first consider the extremal solutions, which describe D-branes
\pol\ at strong coupling. The extremal limit is $r_0 \to 0$, $\alpha \to
\infty$ with $Q$ fixed.  For $n < 4$, the curvature of the extremal
solution in the static frame diverges as $r \to 0$, which is similar
to the dilaton black hole in which the horizon has become singular in
the extremal limit.  For $n=4$ (i.e., the threebrane), the curvature
and the area of the $5$-spheres approach constant values as $r \to
0$. Finally, for $n > 4$, the static frame curvature of the extremal
solution {\it vanishes} as $r \to 0$. This is related to the fact that
at small $r$, $R(r) \sim r^{1-n/4}$, so the area of the $n+1$-spheres
diverges as $r \to 0$.  Since timelike geodesics reach $r=0$ in finite
proper time, this raises the question of whether the spacetime can be
extended beyond this boundary. A smooth extension is known \ght\ for
the threebrane ($n=4$), but not for any of the lower branes.

It is clear from \pbgeoda, \pbgeodb\ that no such extension is
possible.  For all $n \neq 4$, the tidal forces diverge like $1/r^2$
as $r \to 0$. Thus, the solutions with $n >4$, which seem non-singular
from the point of view of the static frame curvature, are actually
singular. (This could also have been inferred from the fact that the
area of the spheres diverges in a finite proper time along a timelike
geodesic, since this implies that the separation between two such
geodesics diverges in finite proper time, and hence the tidal forces
must diverge along the geodesic.) Since the static frame curvature of
these solutions vanishes as $r \to 0$, the $\alpha'$ corrections
vanish, and so these solutions should still be singular when all
$\alpha'$ corrections are included. This is similar to the singular
gravitational plane wave solutions discussed in \hs. For $n=4$, the
tidal forces remain finite at $r=0$, as required by the existence of a
smooth extension.

Now we consider the non-extreme solutions. For any $n$, the curvature
near the horizon $r=r_0$ in the static frame is small if $r_0^2 \cosh
\alpha \gg 1$. This insures that the area of the $r=r_0$ sphere at
constant $y_i$ is large. The total area of the horizon is proportional
to the volume $V$, so it can be made large by a suitable choice of
$V$. On the other hand, the tidal forces will be large, for $n \neq
4$, if $r_0 \ll 1$. We can satisfy this in conjunction with the other
conditions simply by taking $\alpha$ sufficiently large. As before,
this is corresponds to a near extremal configuration.

\subsec{Solutions with several charges}

As our final example, we consider the ten dimensional solution with
three charges which gives, on dimensional reduction, the
five-dimensional black holes whose entropy
\refs{\ascv,\ghrev,\jmthes} and Hawking radiation
\refs{\dam,\ms} has
been explained in terms of an effective string picture derived from
D-brane calculations.

The string metric for this solution is \refs{\cvyo,\hms}
\eqn\tcmet{\eqalign{ds^2 = &f_1^{-1/2} f_5^{-1/2} \left[ -dt^2 +
dx_5^2 + {r_0^2 \over r^2} (\cosh \sigma dt + \sinh \sigma dx_5)^2
\right] + f_1^{1/2} f_5^{-1/2} dx_i dx^i \cr &+ f_1^{1/2} f_5^{1/2}
\left[ \left( 1 - {r_0^2 \over r^2} \right)^{-1} dr^2 + r^2 d\Omega_3
\right],}} 
where $i = 6,7,8,9$, 
\eqn\tcfs{f_1(r) = 1 + {r_1^2 \over r^2}, \qquad f_5(r) = 1 + {r_5^2 \over
r^2},}  
and $r_1 = r_0 \sinh \alpha$, $r_5 = r_0 \sinh \gamma$. The dilaton is
$e^{2 \phi} = f_1/f_5$, and the integer charges are
\eqn\qone{Q_1 = {V r_0^2 \over 2g} \sinh 2 \alpha,}
\eqn\qfive{Q_5 = {r_0^2 \over 2g} \sinh 2 \gamma,}
and
\eqn\nch{n = {R^2 V r_0^2 \over 2g^2} \sinh 2 \sigma,}
where the volume in the $6789$ directions is $(2\pi)^4V$, and the
radius of the $5$ direction is $R$. The mass is 
\eqn\tce{M = {RV r_0^2 \over 2g^2} (\cosh 2\alpha + \cosh 2 \gamma +
\cosh 2 \sigma).}
We will also set $r_n = r_0 \sinh \sigma$. It is also convenient to
define  
\eqn\tcf{F(r) = f_1^{-1/2} f_5^{-1/2}
\left(1-{ r_0^2 \over r^2} \right),}
\eqn\tcg{G(r) = f_n = 1+ {r_n^2 \over r^2}, }
\eqn\tcr{R(r) = f_1^{1/4} f_5^{1/4} r {\rm\ and\ } H(r) = f_1^{1/4}
f_5^{-1/4}.} 

This metric is not of the form \gmet, because of the off-diagonal
terms between $t$ and $x_5$; however, it is relatively straightforward
to generalize the analysis given for \gmet\ to this case. We pick an
orthonormal frame 
\eqn\tcframe{\eqalign{(e_0)_\mu &= -f_1^{-1/4} f_5^{-1/4} f_n^{-1/2}
\left( 1- {r_0^2 \over r^2} \right)^{1/2} \p_\mu t, \cr (e_1)_\mu & =
f_1^{1/4} f_5^{1/4} \left( 1 - {r_0^2 \over r^2} \right)^{-1/2} \p_\mu r,
\cr (e_2)_\mu &= f_1^{1/4} f_5^{1/4} r \ \p_\mu \theta_1, \quad (e_3)_\mu =
f_1^{1/4} f_5^{1/4} r \sin \theta_1\ \p_\mu \theta_2, \cr (e_4)_\mu &=
f_1^{1/4} f_5^{1/4} r \sin \theta_1 \sin \theta_2\ \p_\mu \phi, \cr
(e_5)_\mu &= f_1^{-1/4} f_5^{-1/4} \left[ f_n^{-1/2} {r_0^2 \over r^2}
\cosh \sigma \sinh \sigma \ \p_\mu t + f_n^{1/2}\ \p_\mu x_5 \right], \cr
(e_i)_\mu &= f_1^{1/4} f_5^{-1/4}\  \p_\mu x_i,}}
where $i = 6, \ldots, 9$.  The non-vanishing components of the
curvature in this orthonormal frame are $R_{0101}$, $R_{0k0k}$,
$R_{0151}$, $R_{0k5k}$, $R_{1k1k}$, and $R_{klkl}$, where $k=2,\ldots
9$.  Since $R_{0k1k}=0$, this frame will still minimize the curvature
components under radial boosts.

The timelike geodesics in this metric have constants of motion $E$,
$p_5$, $p_i$, and $p_\phi$.  We can choose coordinates on the
three-sphere so that the geodesic lies in the equatorial plane, so the
tangent vector to the geodesics is
\eqn\tcua{u_\mu = \left( -E, { \dot{r} \over F(r)}, 0,0,
p_\phi, p_5, p_i \right).}
Once again, we set $p_5=p_i=p_\phi = 0$. Then from 
the normalization condition, 
\eqn\tcrdot{\dot{r}^2 = E^2 G(r) - F(r).}
It is still true that $R_{0'a0'a} = -\ddot{R}/R$ and $R_{0'i0'i} =
-\ddot{H}/H$ where $a=2,3,4$ and $i=6,\ldots, 9$.
Near $r=r_0$, it is more convenient to write $R
= \tilde{f}_1^{1/4} \tilde{f}_5^{1/4} \sqrt{r_1 r_5}$ and $H =
\tilde{f}_1^{1/4} \tilde{f}_5^{-1/4} \sqrt{r_1/ r_5}$, where 
\eqn\tctfs{\tilde{f}_1 = 1 + { r^2 \over r_1^2},\qquad \tilde{f}_5 =
1 + {r^2 \over r_5^2}.} 
Thus,
\eqn\tcgeod{\eqalign{R_{0'a0'a} &= -{E^2 \over R}(G R'' + G' R'/2) +
\ldots \cr &= -{E^2 \over 16 \tilde{f}_1^2 \tilde{f}_5^2} \left[ f_n (4
\tilde{f}_1'' \tilde{f}_1 \tilde{f}_5^2 - 3 \tilde{f}_1'^2
\tilde{f}_5^2  +  \tilde{f}_1' \tilde{f}_5' \tilde{f}_1\tilde{f}_5) +
2 f_n' \tilde{f}_1'\tilde{f}_1 \tilde{f}_5^2 + (1 \leftrightarrow 5)
\right] + \ldots}} 
Similarly, 
\eqn\tcgeodb{R_{0'i0'i} = -{E^2 \over H}(G H'' + G' H'/2) +
\ldots}
$$= -{E^2 \over 16 \tilde{f}_1^2 \tilde{f}_5^2} \left[ 4 f_n
\tilde{f}_1'' \tilde{f}_1 \tilde{f}_5^2 + 2 f_n'
\tilde{f}_1'\tilde{f}_1 \tilde{f}_5^2 - (1 \leftrightarrow 5) - f_n (2
\tilde{f}_1' \tilde{f}_5' \tilde{f}_1\tilde{f}_5 + 3 \tilde{f}_1'^2
\tilde{f}_5^2 - 5 \tilde{f}_5'^2 \tilde{f}_1^2) \right] + \ldots $$
where, as before, we have included only the terms proportional to $E^2$.

If we assume $r_1, r_5 \gg r_0$, the curvature at the horizon in the
static frame is of order $1/(r_1 r_5)$, so we must take $r_1 r_5 \gg
1$ to make this small. If $r_0, r_n \ll r_1, r_5$, then $\tilde{f}_1,
\tilde{f}_5 \approx 1$ at $r=r_0$, and the tidal forces are then not
large. However, if $r_0, r_n, r_1 \ll r_5$, then the tidal forces are
proportional to 
$ E^2 / r_1^2$, and so can be large if we take $r_1 \ll
1$. If we also take $r_5 \gg 1/r_1$, we can keep the curvature in the
static frame small at the same time. This limit is physically
reasonable, as we can suppose that $V$ is sufficiently large to make
$Q_1$ and $n$ large despite the fact that $r_1, r_n \ll 1$. Many of the
recent calculations of near extremal black hole entropy and emission 
rates in string
theory have assumed $r_0, r_n \ll r_1, r_5$.  However, the case $r_0,
r_n, r_1 \ll r_5$ has been considered in e.g. \refs{\mald,\km}.

\newsec{Discussion}

We have seen that objects falling into large black holes can
experience large tidal forces outside the horizon.  This seems to be a
property of most of the recently discussed near extremal black hole
and black p-brane solutions in string theory. The entropy of all of
the solutions in the previous section has recently been understood in
terms of a correspondence principle \hopo\ which relates the black
holes at strong coupling to the states of strings and D-branes at weak
coupling.  The transition occurs when the $\alpha'$ corrections to the
metric become significant. We have seen that this occurs when the size
of the horizon is of order the string scale (so the static curvature
is of order the string scale) and not when the tidal forces are
large. Thus the agreement found in \hopo\ between the counting of
string states and black hole entropy is not affected by the results
found here.

What are the physical effects of the large curvature outside the
horizon?  Consider a solid object falling toward the black hole. Tidal
forces in the radial direction remain small, but those in the
transverse directions become very large. When they become greater than
the internal pressures and stresses can support, each particle
essentially follows a geodesic.  The object then shrinks to a fraction
of its initial size before it crosses the horizon. In effect, the
object is crushed by the gravitational field outside the black hole.

As mentioned in the introduction, the existence of large curvatures
outside the horizon weakens the spirit of cosmic censorship.  In light
of this, it is natural to ask whether one could form these black holes
from the collapse of charged matter with low density. Recall that
neutral matter forms a Schwarzschild black hole when its average
density is of order $1/M^2$, which is small for large $M$. Naked black
holes typically have a horizon size which is much smaller than the
Schwarzschild solution of the same mass.  Nevertheless, it is still
possible to form some of these black holes at low density. Consider
the dilaton black holes discussed in section {\it 3.1}.  In the near
extremal limit, $M = r_+/(1+a^2)$, so if this mass were confined to a
radius of order the horizon size $R=r_+ \e^{a^2/(1+a^2)}$ it would
have a density
\eqn\den{\rho \sim {r_+ \over r_+^3 \e^{3a^2/(1+a^2)}} =
{\e^{(2-a^2)/(1+a^2)} \over (r_+ \e)^2  }.} 
The tidal forces on infalling observers are of order $1/(r_+ \e)^2$,
so for $a^2<2$, the parameters can be chosen so that the mass forms a
black hole (with large tidal forces)
when its average density is arbitrarily small. 
This includes the value $a=1$ which is important for string theory.

The existence of large tidal forces outside the horizon could play a
role in resolving the black hole information puzzle for these black
holes, since the large tidal forces may provide a mechanism for
transferring information from the ingoing matter to the outgoing
Hawking radiation. To examine this further, let us consider the
calculation of Hawking radiation in one of these backgrounds.  In the
simplest case, one studies the propagation of a free scalar field in
the black hole background, and computes the mixing of positive and
negative frequency modes at past and future null infinity. Since the
scalar wave equation can be evaluated in any frame, this equation does
not see the large tidal forces. In other words, the effective radial
potentials have a size set by the curvature in the static frame, which
remains small.\foot{This is different from the case studied in
\holwil\ where it was shown that these potentials can become large as
the horizon shrinks to zero size for certain dilatonic black holes.}

However, if we consider instead a test string falling into one of
these black holes, the large tidal forces will cause it to become
excited. So even if the string starts in a massless state at infinity,
before it falls through the horizon, there is a high probability that
it will be excited into a very massive state.  This is similar to the
effect on strings propagating through strong gravitational plane waves
\hs. In fact, the physics is very similar, since the curvature is
becoming large and almost null near the horizon.

Proponents of unitarity have been looking for loopholes in Hawking's
original arguments that information must be lost when black holes
evaporate. The fact that the usual semi-classical calculations do not
see the large tidal forces, while test strings do, may be an important
clue to what is missing.

Even if these large tidal forces prove to be important for the
information loss problem for these black holes, one still has to worry
about black holes which are far from extremality, where the tidal
forces remain small outside the horizon. It should be noted that even
for Schwarzschild, one expects that string $\alpha'$ and quantum
corrections will destroy the boost invariance of the curvature. Thus
generically, objects falling into a black hole feel different tidal
forces depending on their energy. The tidal forces for Schwarzschild
should remain small when the conserved energy $E$ is of order one.
However, one might speculate that some quantum analogs of this
classical effect might resolve the information puzzle.  Some
potentially related mechanisms, which also depend on the difference
between static and infalling observers, have been discussed in the
context of string theory by Susskind and others \sussug.

\vskip 1cm

{\bf Acknowledgments}

\vskip .5cm

We wish to thank S. Giddings and H. Yang for helpful comments.
This research was supported in part by NSF Grant PHY95-07065. SFR also
acknowledges support from the Natural Sciences and Engineering
Research Council of Canada. 

\listrefs
\end